# Digital Urban Sensing: A Multi-layered Approach

Enwei Zhu[1,2], Maham Khan[1], Philipp Kats[1,3], Shreya Santosh Bamne[1], Stanislav Sobolevsky[1,4*]

[1]Center for Urban Science and Progress, New York University, New York, United States of America

[2]Hang Lung Center for Real Estate, Tsinghua University, Beijing, China

[3]Mathematical modeling of non-equilibrium processes in oil and gas production laboratory, Kazan Federal University, Russia

[4] Institute Of Design And Urbanism of Saint-Petersburg National Research University Of Information Technology, Mechanics And Optics (ITMO), Russia

* Email: sobolevsky@nyu.edu

## Abstract

Studies of human mobility increasingly rely on digital sensing, the large-scale recording of human activity facilitated by digital technologies. Questions of variability and population representativity, however, in patterns seen from these sources, remain major challenges for interpreting any outcomes gleaned from these records. The present research explores these questions by providing a comparison of the spatial and temporal activity distributions seen from taxi, subway and Citi Bike trips, mobile app records, geo-tagged Twitter data as well as 311 service requests in the five boroughs of New York City. The comparison reveals substantially different spatial and temporal patterns amongst these datasets, emphasizing limitations in the capacity of individual datasets to represent urban dynamics in their entirety. We further provide interpretations on these differences by decomposing the spatial distributions with working-residential balance and different propensities for demographic groups to use different activities. Nevertheless, the differences also highlight the opportunity to leverage the plurality to create multi-layered models of urban dynamics. We demonstrate the capacity of such models to advance urban zoning and socio-economic modeling – two common applications of digital urban sensing.

# 1 Introduction

Understanding human mobility patterns can be of critical importance for urban planning, in sectors as diverse as transport, emergency management, epidemiology, land management, water and waste planning, only to name a few. And whilst urban mobility is not observable in its entire complexity, an increasing number of studies rely on various *proxies* to mobility, as represented by different types of available data, moving beyond static data such as Census and LEHD Origin-Destination Employment Statistics (LODES) data, towards more dynamic geospatial data representing various aspects of human activity. This gives rise to the new concept of a "digital census" – an opportunity to measure the presence of populations in real time.

Recently, this notion of a digital census has seen an increasing number of urban applications. Cell phone connections [1-8], credit card transactions [9-11], GPS readings [12, 13], geo-tagged Twitter [14-17] and Flickr [18], 311 service requests [19], Wi-Fi [20] and Bluetooth [21] connections have all been demonstrated to be useful proxies for urban dynamics.

A key challenge for digital census is the limited representativeness of the currently available urban data–no data layer representing a single aspect of human activity can fully represent the dynamics of urbanism. Not everyone uses social media or mobile services, no app is frequented by all inhabitants of a city (and even if used occasionally by a person it is almost never used at all times), only a limited slice of urban commuters is facilitated by any one transport service, etc. This limits the utility of any individual dataset, as well as the credibility of any findings that may be derived from it.

In cities such as New York, however, a critical mass of datasets has already been accumulated to warrant a comprehensive comparative study of the ways in which human mobility can be represented. This allows an investigation of the extent to which each specific dataset can be used as an *unbiased* proxy for urban mobility and any biases that may need to be acknowledged. One recent paper performed a similar study for the city of Singapore comparing taxi and public transit data [22].

Two common applications of such digital proxies to human mobility are 1) to represent real-time distributions of urban populations at a *specific* moment or typical period of time, and 2) to assess the dynamics of urban populations in a specific location over a *period* of time. While the first is important for long-term planning of urban infrastructure (specifically, transportation demand), the second, besides adding a component on dynamism to such plans, is commonly used as a *characteristic* of urban location, e.g. for land use classification applications [23, 24], zoning or socio-economic modeling [19].

In the present study, we first compare the spatial distributions and dynamics of urban activity as observed from taxi, subway and Citi Bike trips; mobile app records; geo-tagged Twitter as well as 311 service requests. We find that, while exhibiting some crucial points of similarity, each dataset provides *unique* insights into urban dynamics as well as potential intrinsic biases. We then try to provide some interpretations on the differences between these datasets. Specifically, we decompose the spatial distributions with working-residential balance and different propensities for demographic groups to use different activities. These results suggest a multi-layered modeling approach which benefits from the diversity of information contained in each of the data layers. Last, we demonstrate the superiority of such modeling approach by two typical applications of digital urban sensing – urban zoning and socio-economic modeling.

## 2 Data and pre-processing

The following datasets were utilized in this study:

1. 311 (service requests): https://nycopendata.socrata.com/Social-Services/311-Service-Requests-from-2010-to-Present/erm2-nwe9

2. Mobile (cellular device data): http://skyhook.carto.com/

3. Twitter activity (feed of tweets with geo-tags collected via API): www.twitter.com

4. Taxi and Uber (pick-ups and drop-offs): nyc.gov/html/tlc/html/about/trip_record_data.shtml

5. Subway (entries and exits): web.mta.info/developers/turnstile.html

6. Citi Bike (individual trip details which include start and end locations and timestamps): https://www.citibikenyc.com/system-data

The study compares the spatial distributions and dynamics of urban activity during one month (July, 2017), due to constraints on availability of mobile data. We also compare these datasets to the static residential and working population baseline distributions, as observed in census and LODES data, found at https://lehd.ces.census.gov/data. The start- and end-records are separately aggregated for taxi, subway and Citi Bike.

**2.1 Spatial pre-processing**

Each of the datasets (labeled 1-6 above), in addition to census and LODES data, are transformed into a vector representing activity/population counts in each of 263 taxi zones (the finest level of spatial granularity constraining our analysis).

**2.2 Spatio-temporal pre-processing**

Each of the six datasets are transformed into a spatiotemporal matrix, of which the first and second dimensions are consistent with 263 taxi zones and 56 three-hour intervals through a typical business week (eight three-hour bins in a day and seven days in a week). The activity counts in the same three-hour bin are averaged over different weeks in the entire month of July, 2017.

## 3 Comparison of the spatial activity distribution

Figure 1 below visualizes the spatial densities of all six activities, all appearing to represent considerably different mobility patterns. These differences are explored quantitatively in a correlation matrix, as illustrated in Figure 2. The residential population (based on census data) and working population (based on LODES data) distributions are also included for comparison.

Most of the correlation coefficients tend to be positive, suggesting a common tendency shared among all the spatial distributions. More specifically, the spatial distributions of 311 and mobile activities appear to be strongly correlated with the residential population, while the distributions of Twitter, taxi, subway and Citi Bike activities lean towards greater similarity to the working population distribution.

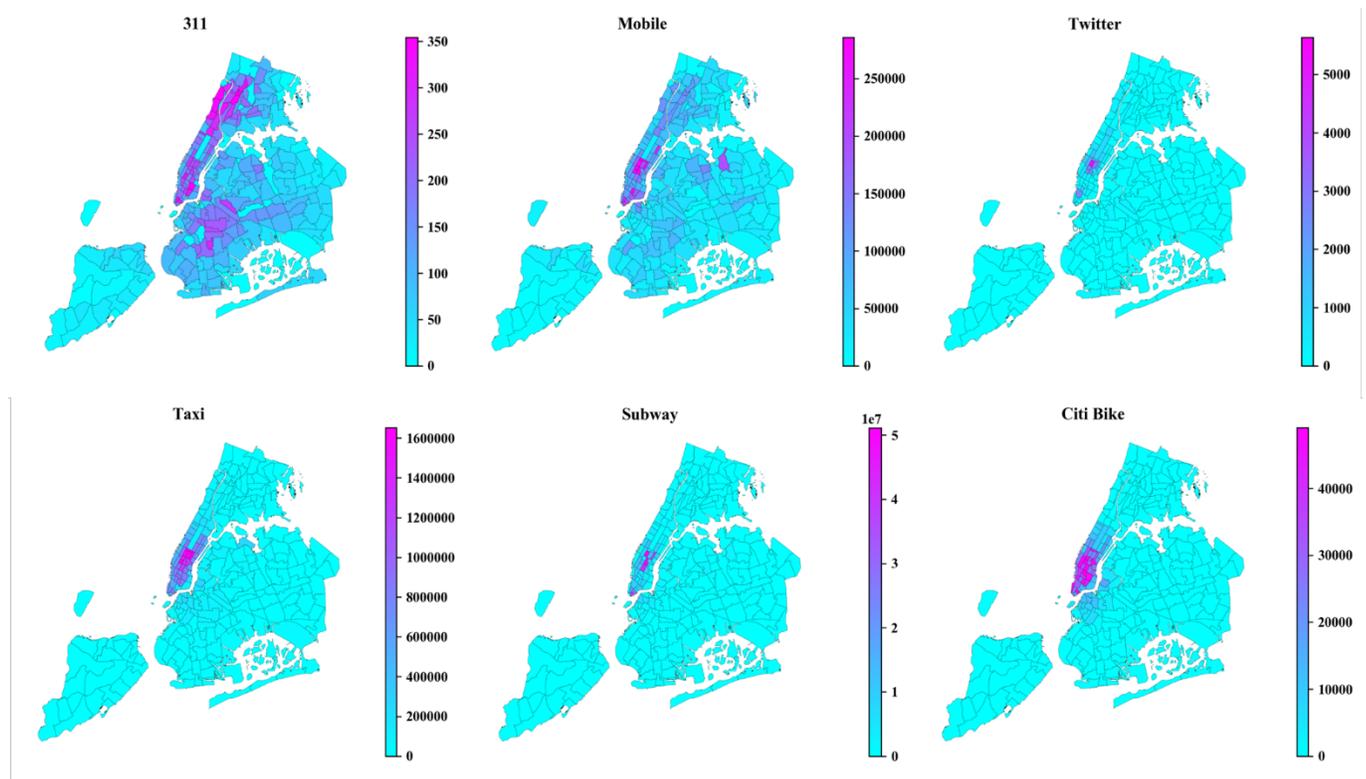

Figure 1 Spatial distribution of activity density for each dataset (records per square kilometer)

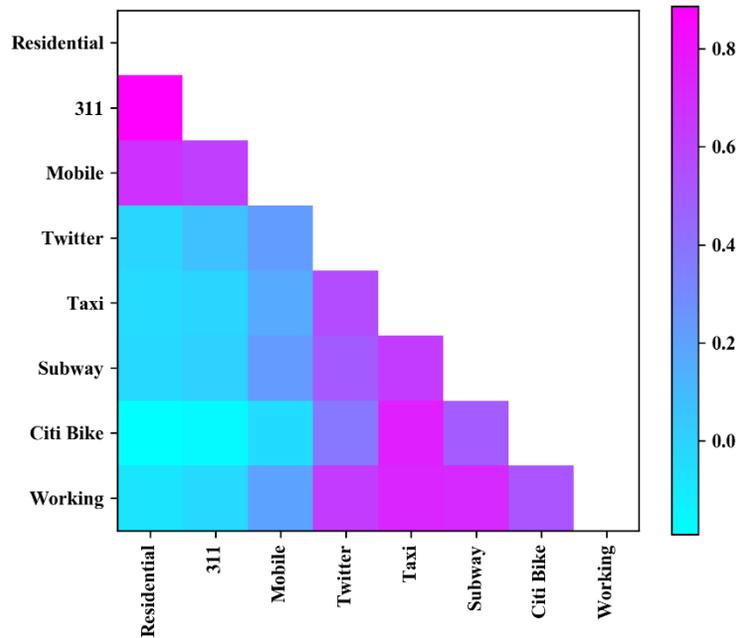

Figure 2 Pearson correlation coefficients between spatial distributions of six types of activities as well as residential and working populations

## 4 Comparison of the temporal activity distribution

In order to compare the temporal patterns of the activities, we aggregate the data to a weekly timeline for each of the six datasets, and then apply L1 normalization to the timelines. Figure 3 presents the normalized timelines, as well as an average baseline plotted in gray for each subplot.

First, each of the six timelines clearly shows seven cycles during the typical week, consistent with our expected cyclical activity patterns that daytime should produce more activities than nighttime. Second, within each timeline, the two days in weekends appear to show different patterns from the weekdays. For example, the subway activity counts on weekends are significantly smaller than those on weekdays. In addition, taxi and Citi Bike (as well as subway, albeit to a lesser degree) show distinct bimodal peaks in the morning and evening commute times for each weekday, while such pattern disappears in the weekends. Third, 311, mobile and Twitter are basically unimodal for each day, although they are slightly out of phase. Specifically, 311 timeline peaks slightly earlier than the baseline, while mobile and Twitter peak slightly later.

Figure 4 displays the correlation matrix comparing the similarities between six normalized activity timelines. All the coefficients are positive, consistent with the common cyclical pattern shared among the six timelines. In addition, higher correlation coefficient indicates more similar weekly patterns. For example, mobile and Twitter show the highest correlation, and their timelines are almost identical during the week—both show unimodal peak slightly lagged behind average baseline for each day. On the other hand, the weakest correlations are observed between Citi Bike and Twitter, and 311 and Twitter, respectively.

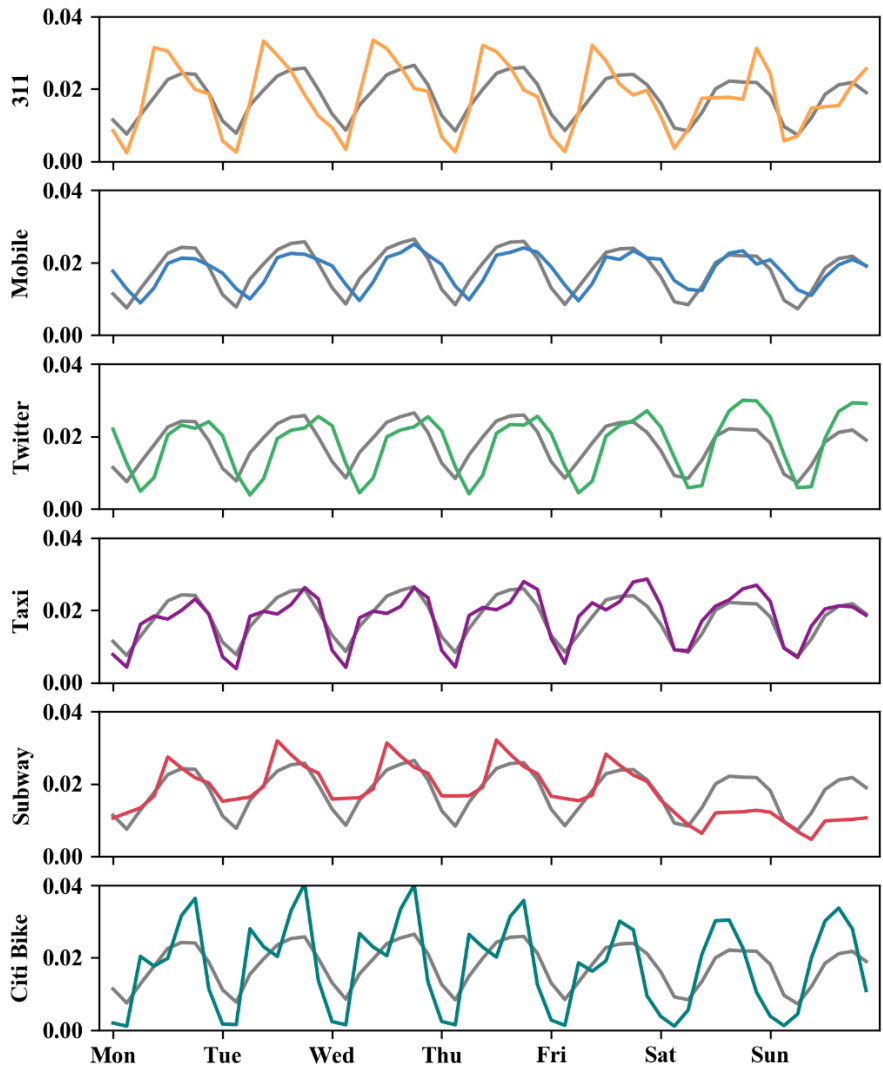

Figure 3 Weekly temporal activity distribution for each of the six datasets (normalized relative activity volume per each 3 hour bin).

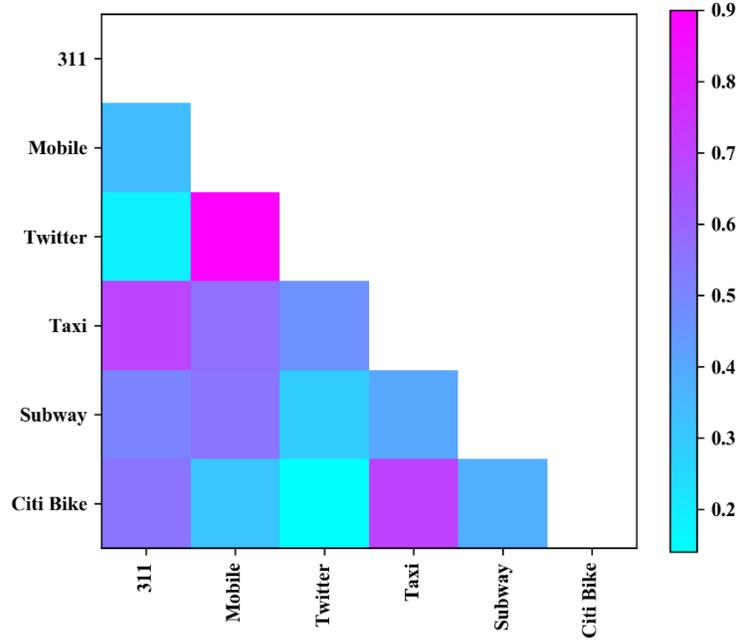

Figure 4 Pearson correlation coefficients between normalized timelines of six types of activities

## 5 Decomposition by working-residential balance and demographic groups

We are particularly interested in explaining why the spatial distributions are different across activities. First, we notice that people can commute between their residential and working places, and they may have different propensities to use certain activity in these two places. In that case, each activity would be representative to a particular balance between working and residential populations. To this end, we assume that the observed spatial distribution of each activity is a linear combination of residential and working populations:

$$A = \alpha \cdot W + (1 - \alpha) \cdot R$$

where $A$, $W$ and $R$ are the normalized vectors indicating the spatial distributions of activity, working population and residential population, respectively; $\alpha$ is a coefficient indicating the balance of activity between working and residential populations; a higher $\alpha$ suggests that the activity is closer to working population. We estimate coefficient $\alpha$ by fitting the linear regression:

$$A - R = \alpha \cdot (W - R)$$

Figure 5 presents the estimated coefficient $\hat{\alpha}$ for each of the six activities. The $\hat{\alpha}$ for the 311 dataset is very close to zero, the lowest of all the datasets, suggesting the spatial distribution of 311 activities is very close to the residential population. This result is plausible as 311 complaints are typically made from residential places. On the other hand, the three transport datasets (taxi, subway and Citi Bike) as well as Twitter activities show estimated $\hat{\alpha}$ greater than 0.7, suggesting that people prefer to use these services at or in relation to their work.

Figure 6 presents the estimated $\hat{\alpha}$ for each three-hour bins over a typical week for the six activities. The overall levels of $\hat{\alpha}$ are consistent with those in Figure 5, while the $\hat{\alpha}$ timelines show clear cyclical patterns during the week. Typically, the timelines peak around the middays, while reach bottom during the hours after midnights. In addition, the $\hat{\alpha}$ values appear to be smaller during weekends than weekdays. These results are exactly consistent with the people's commute patterns.

In summary, the datasets differ substantially in the extents to which they tend towards residential or working populations, which are further dynamic over time. We argue that this is one of the most important factors accounting for the differences in the spatial distributions of these activities.

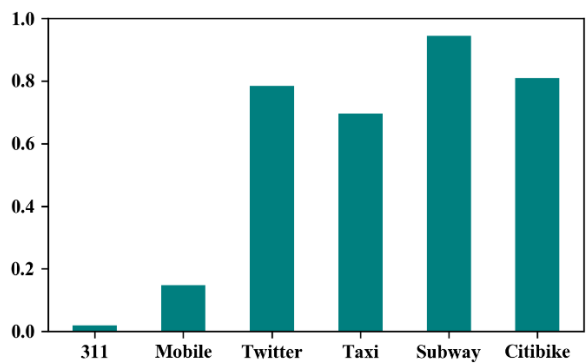

Figure 5 Balances between residential and working population contributions for the six activities (1 – all activity comes from working population, 0 – all from residential).

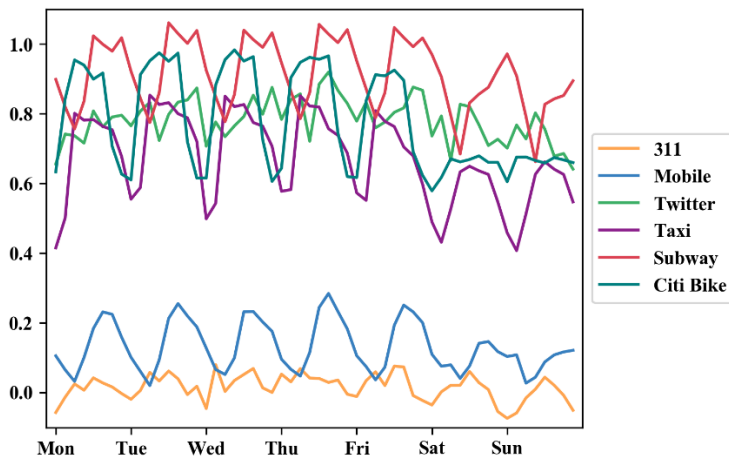

Figure 6 Balances between residential and working populations over a typical week

We then include more potential information to explain the differences between the six activities. Specifically, we hypothesize that these activities comprise different components over demographic groups (genders, ages, races and education levels); in other words, people in particular demographic group have different propensities to use different activities. First, we use the $\hat{\alpha}$ estimated above to extrapolate between *W* and *R* for each demographic group, producing the spatial structure of "acting" population related to the activity:

$$P_G = \hat{\alpha} \cdot W_G + (1 - \hat{\alpha}) \cdot R_G$$

where $P_G$, $W_G$ and $R_G$ are the acting, working and residential populations for demographic group G, respectively. Last, we regress *A* on the resulting demographic groups:

$$A = \sum_G \beta_G \cdot P_G$$

where $\beta_G$ represents the propensity for demographic group *G* to use the activity. However, when estimating the above equation, we find that $P_G$ on different groups are highly correlated, resulting in unreasonable values for $\hat{\beta}_G$. Hence, we assume a prior distribution centered at average propensity for $\beta_G$, so impose L2-regularization to avoid $\beta_G$ diverging too away from average propensity.[1] In other words, we minimize:

$$\left\| A - \sum_G \beta_G \cdot P_G \right\|_2^2 + \lambda \cdot \sum_G \left( \beta_G - \frac{1}{N} \right)^2$$

where $\lambda$ is the L2-penalty term, and *N* is the number of demographic groups.

Figure 7 presents the estimated propensities over different demographic groups for the six activities. For each kind of demographic groups, the propensity distributions show significant differences across different activities. It implies that these activities are popular in different demographic groups, which may also help explain the differences in the spatial distributions of the activities. In addition, most of the estimated propensities are plausible and interpretable. For example, "15 to 29" and "30 to 59" age groups are the main source of work commute, so they show substantially greater propensities than other two age groups. Besides, people with "bachelor or higher" education levels are more likely to use Twitter, taxi and Citi Bike services.

---

[1] L2-regularization may result in biased coefficient estimation. However, our objective here is to show the difference in propensities across activities, so the difference between coefficients, rather than the absolute coefficient values, are of our interest.

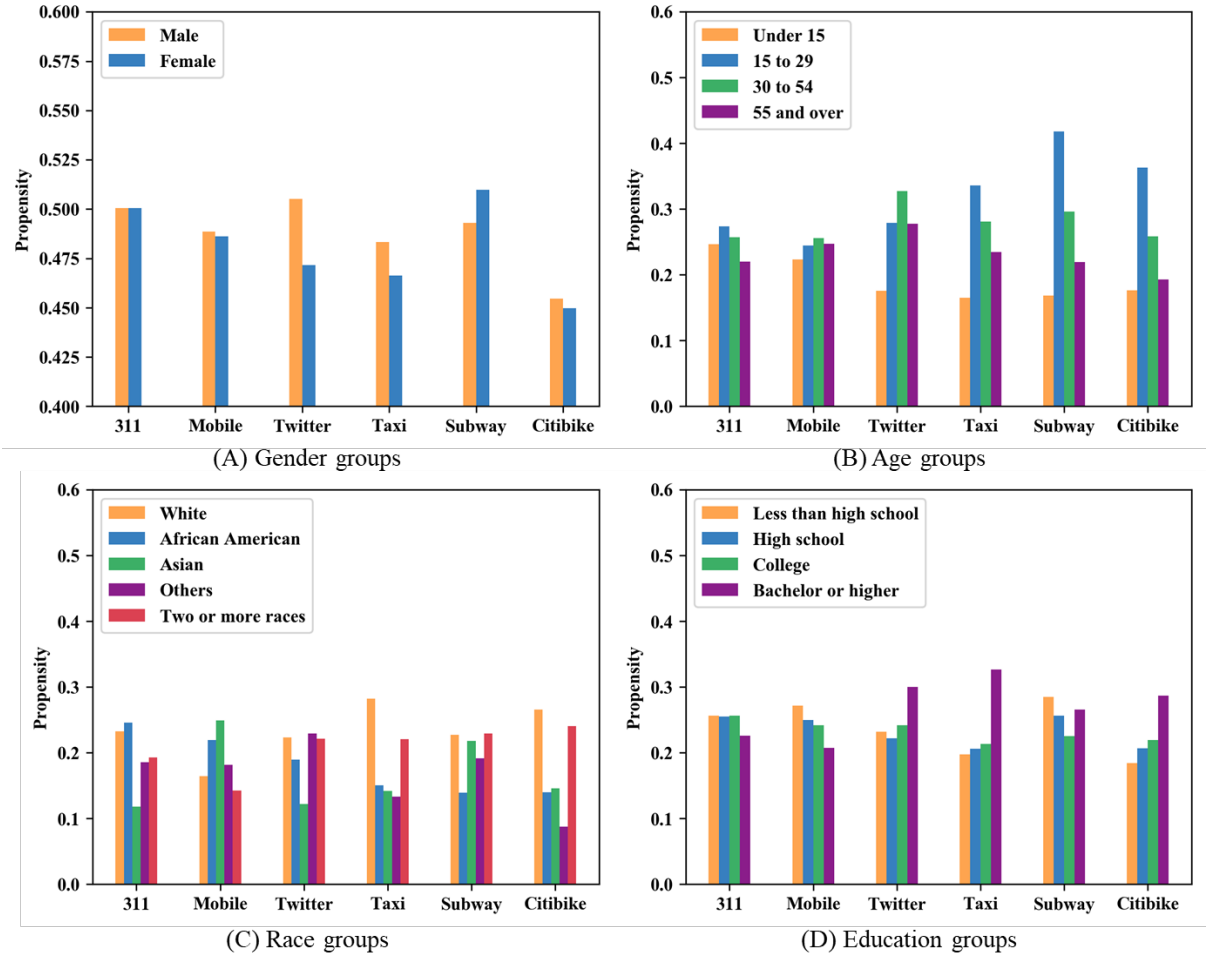

Figure 7 Propensities for different demographic groups to perform the six types of activities

## 6 Urban zoning based on the temporal signatures of human activity

Weekly timelines of human activity were often used to characterize the context of the area, including land use classification applications [23, 24]. In this section we leverage all six weekly activity timelines above to group areas with similar behavior profile across the city. The premise here is to outline areas of similar socio-economic context and role in urban environment. While different types of activity reveal different spatio-temporal patterns as seen from the sections above, we believe that considering those different activities together might further enrich our understanding of the local context.

In our clustering analysis, the features are the timelines of relative activity frequencies. Specifically, we use the activity counts within each of the three-hour bins of a typical week (see section 2.2), and undertake L1 normalization of the activity count vector, so there are 56 features for each dataset and 336 features altogether.

For locations with insufficient activity counts, spatial smoothing is introduced, which aggregates and averages over neighboring data points. In case the local data volume remains insufficient after smoothing, these locations are omitted. We use k-means as the clustering method. In order to select an optimal number of clusters we utilize a standard Elbow method, comparing within-cluster sum of squares. According our experiments, k=5 seems to be the best choice, and for consistency it is applied to all datasets.

Figure 8 plots the clustering areas identified by each of the six datasets, as well as the corresponding activity timelines for each cluster. The resulting clusters have a clear spatial pattern – revealing tourism-intensive zones, low-density residential areas, Central Business District, etc. Many clusters are reasonably consistent through different data sources. Still, each dataset, due to the nature of each kind of activities, provides a unique angle on city partitioning.

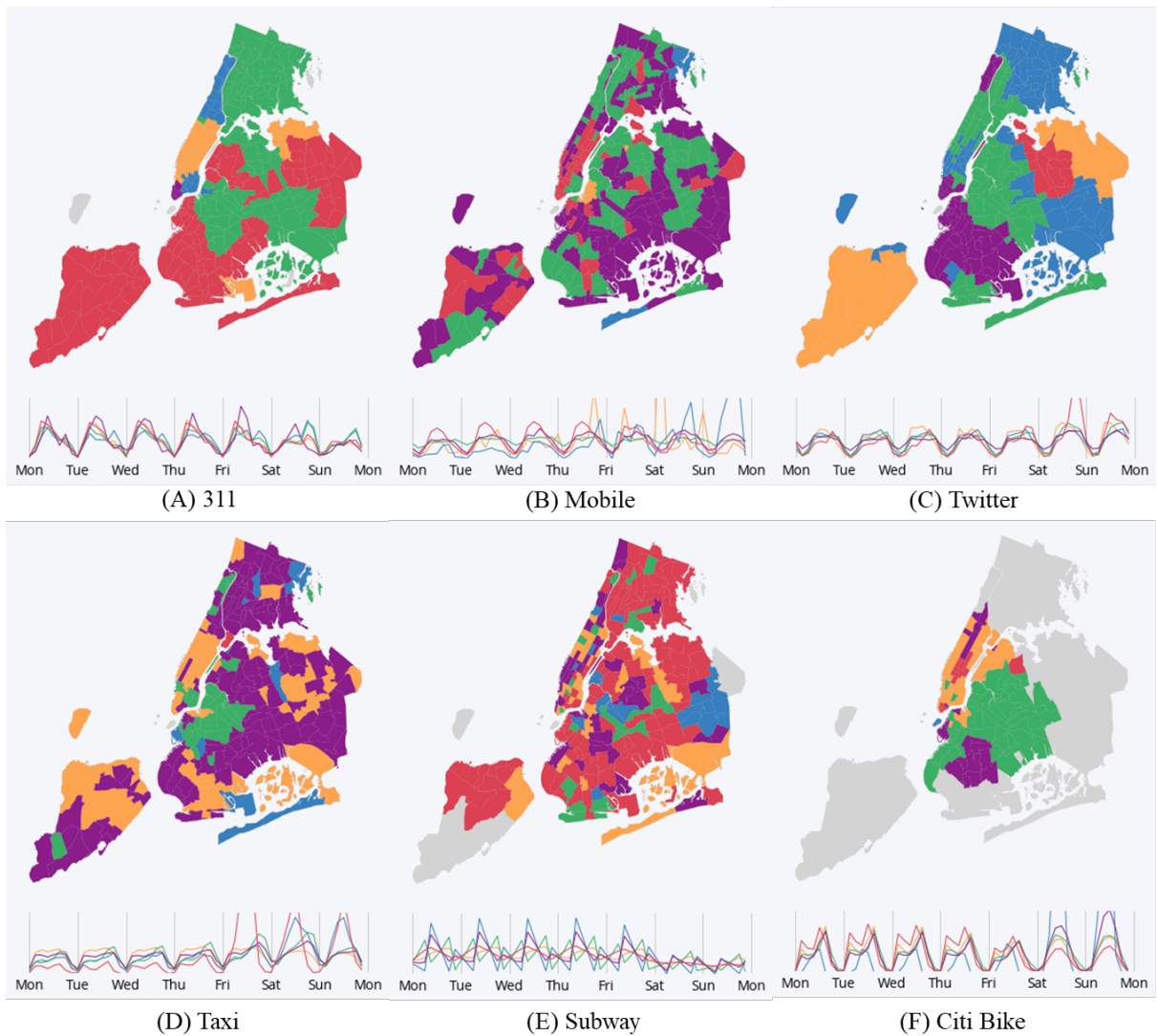

Figure 8 Clustering results and corresponding relative activity timelines for each cluster on each of the six datasets

Figure 9 illustrates the clustering results using all the six activities together. The resulting sets appear to be more spatially cohesive, since they draw upon the insights from all the types of activities considered. As time series pick different patterns of activities through the city, corresponding clusters represent areas of different socio-economic profile, as presented on Figure 10. Each cluster has a distinctive set of characteristics.

In Figure 11, we calculate adjusted Rand metric (ARM) [25] and the Silhouette score [26], to estimate the quality of the partitions over socio-economic properties. In the first case, we compare partitions to quintiles of demographic features, treated as categories. Each partition is then characterized by the average score over all socio-economic properties. We further compute Silhouette scores of the cluster over all features treated as single multidimensional space. As a result, 311 complaints-based partition performs the best both in ARM, with the overall partition on the second place. The Overall partition, however, performs the best in the silhouette score, suggesting that partition based on all data allows us to better fit clusters of areas with similar combinations of socio-economic properties.

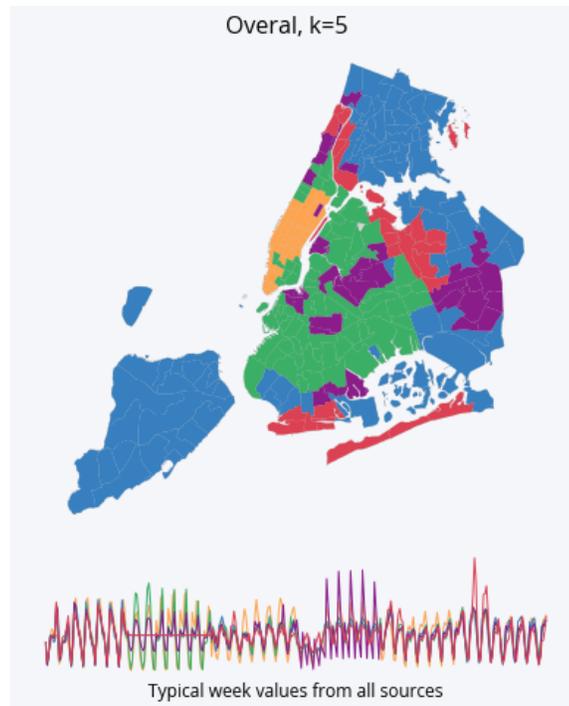

Figure 9 Clustering results and corresponding relative activity timelines for each cluster on all six datasets

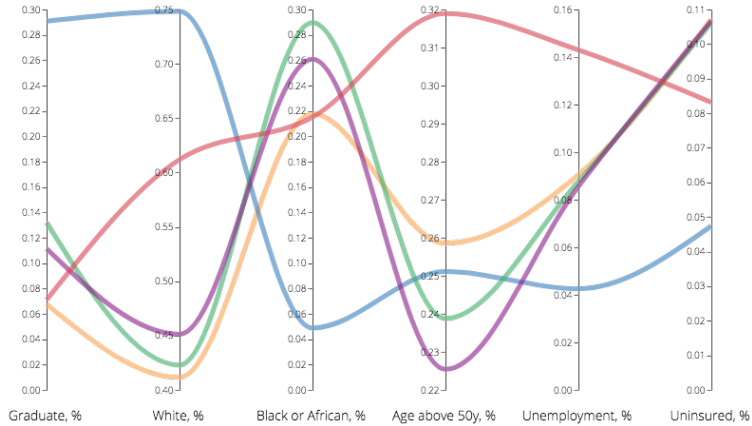
Figure 10 Demographic properties of the clusters identified by data from all six datasets together.

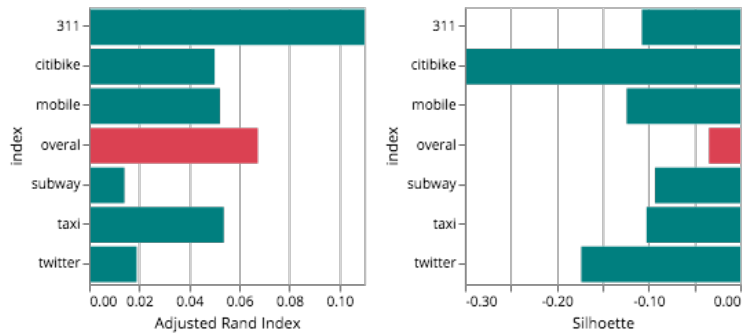
Figure 11 Partition metrics (Adjusted Rand Index and Silhouette) highlighting utility of the overall partition

## 7 Leveraging temporal signatures of human activity for socio-economic modeling

Following the clustering analysis, we explore whether the temporal signatures of different datasets can be used to model socio-economic quantities. The approach to modeling socio-economic variables with one single type of mobility activity, such as 311 service requests [19], Twitter [17] or credit card activity [9-11] has previously proven to be useful. Presently, whilst each dataset seems to make a unique contribution to mobility modeling, we are interested in whether combining all the datasets together can substantially improve predictive power.

Similar to clustering analysis, the features are the activity timelines. The target variables include the percentages of White, African-American and Asian populations, the percentages of residents with bachelor's and graduate degrees, the median income of households, the percentage of residents below poverty, unemployment rates, the percentage of residents uninsured, as well as median house prices per square feet.

We employ three kinds of models: Lasso Regression [27], Random Forest Regression [28, 29] and Gradient Boosting Regression Trees [30, 31]. For each model, we specify different sets of hyper-parameters and treat each set as an independent model. Specifically, for Lasso Regression, we use a penalization alpha values of 1e-5, 3.3e-5, 1e-4, 3.3e-4, …, 1. For Random Forest and Gradient Boosting Regression Trees, we use tree numbers of 5, 10, 20, 50, 100, 200, 500 and 1000, maximum depths of trees for 3, 5 and "fully expanded", and try "the number of features considered to expand a node" for the number of all features and its square root and its logarithm. In total, we have 155 different models.

We use a five-fold cross-validation procedure to calculate the out-of-sample R-squared for each model. We repeat this procedure for 20 times, and then calculate the average R-squared values as well as corresponding standard deviations over the 20 cross validations. Typically, Gradient Boosting Trees outperform the other two kinds of models. Finally, we select the model (and hyper-parameters) with the best performance for each dataset and target variable.

Figure 12 compares the best average R-squared values over the different datasets for each socio-economic variable. It illustrates that performances are substantially different across datasets. Specifically, the timelines of 311 and Taxi datasets can achieve the highest R-squared values, while the mobile dataset usually produces the lowest. This results suggests differences in the capacities of these datasets to reflect local socio-economic conditions. Further, such differences show heterogeneity across different socio-economic variables. For example, the 311 and taxi datasets appear to show advantages in modeling different socio-economic variables.

Table 1 outlines the R-squared values resulting from combining all datasets and the improvement achieved in comparison with the highest R-squared from any single dataset. All such differences are positive, suggesting that combining all the datasets always gives better modeling performance than any single dataset alone can provide. In addition, the improvements are typically larger than or at least similar in magnitude to the corresponding standard deviations, suggesting that those improvements are substantial.

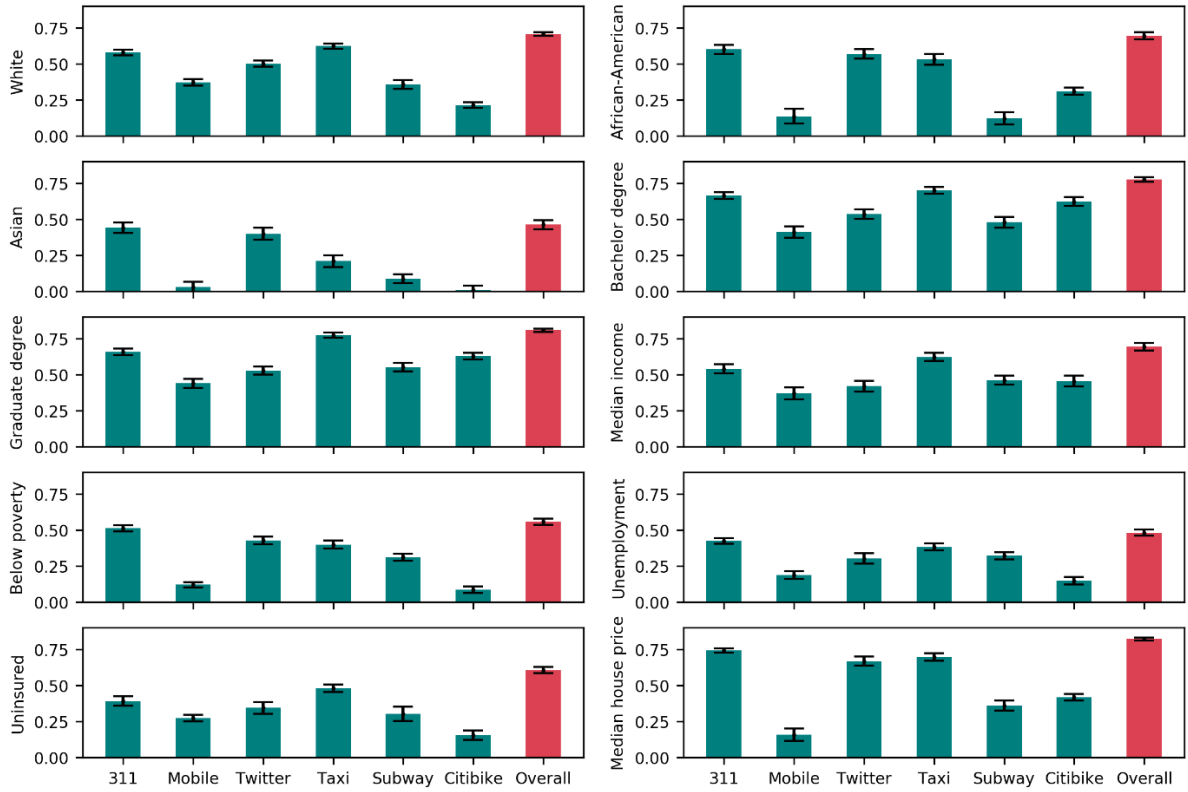

Figure 12 Best out-of-sample average R-squared values for modeling socio-economic variables with different datasets (Error bars represent standard deviations).

| Socio-economic variable | R-squared (standard deviations) | Improvement of R-squared |
|---|---|---|
| White | 0.707 (0.014) | 0.084 |
| African-American | 0.695 (0.024) | 0.095 |
| Asian | 0.462 (0.031) | 0.021 |
| Bachelor degree | 0.775 (0.016) | 0.073 |
| Graduate degree | 0.808 (0.012) | 0.034 |
| Median income | 0.693 (0.025) | 0.071 |
| Below poverty | 0.558 (0.021) | 0.047 |
| Unemployment | 0.481 (0.022) | 0.056 |
| Uninsured | 0.607 (0.021) | 0.126 |
| Median house price | 0.815 (0.011) | 0.067 |

Table 1 Best out-of-sample average R-squared values for modeling socio-economic variables with all datasets and improvements against the best result obtained from single datasets.

## Conclusions

Datasets recording various aspects of human activity are often considered viable and representative proxies to urban dynamics. In the present paper we demonstrate that such datasets from New York City, such as taxi, subway and Citi Bike trips, mobile app records, Twitter and 311 service requests, provide valuable but quite different spatio-temporal insights. This makes it difficult to claim any one of such datasets as a sufficient *standalone* proxy. However, each dataset contributes its unique perspective, which when combined together can provide a much more comprehensive picture of urban dynamics. We show that a multi-modal combination of the six datasets substantially improves the possibility of learning important local socio-economic patterns and modeling socio-economic quantities, as compared to using each dataset alone, emphasizing importance of a multi-layered perspective to urban digital sensing.